# Observation of rotation-induced light localization in waveguide arrays

Chunyan Li,[1,2,*] Antonina A. Arkhipova,[2,3,*] Yaroslav V. Kartashov,[2] Sergey A. Zhuravitskii,[2,4] Nikolay N. Skryabin,[2,4] Ivan V. Dyakonov,[4] Alexander A. Kalinkin,[2,4] Sergey P. Kulik,[4] Victor O. Kompanets,[2] Sergey V. Chekalin,[2] and Victor N. Zadkov[2,3]

[1]School of Physics, Xidian University, Xi'an, 710071, China
[2]Institute of Spectroscopy, Russian Academy of Sciences, 108840, Troitsk, Moscow, Russia
[3]Faculty of Physics, Higher School of Economics, 105066 Moscow, Russia
[4]Quantum Technology Centre, Faculty of Physics, M. V. Lomonosov Moscow State University, 119991, Moscow, Russia
*Corresponding author: chunyanli@xidian.edu.cn

**ABSTRACT:** We study both, experimentally and theoretically, propagation of light in the fs-laser written rotating square waveguide arrays and present the first experimental evidence of light localization induced by the rotation of periodic structure in the direction of light propagation. Such linear light localization occurs either in the corners of truncated square array, where it results from the interplay between the centrifugal effect and total internal reflection at the borders of truncated array, or in the center of array, where rotation creates effective attractive optical potential. The degree of localization of linear bulk and corner modes emerging due to the rotation increases with the increase of rotation frequency. Consequently, corner and bulk solitons in rotating waveguide arrays become thresholdless for sufficiently large rotation frequencies, in contrast to solitons in non-rotating arrays that exist only above power threshold. Focusing nonlinearity enhances localization degree of corner modes, but surprising initially it leads to broadening of bulk nonlinear states, followed by their re-localization at high input powers. Our results open new prospects for control of evolution of nonlinear multidimensional excitations by dynamically varying potentials.

## INTRODUCTION

Periodic potentials, whose linear spectra are characterized by the presence of allowed and forbidden bands, profoundly affect evolution of linear excitations, as demonstrated in diverse areas of physics including photonics, acoustics, and matter wave physics. Expansion rate, velocity, and structure of linear wavepackets in such potentials depend on their momentum and dispersive properties of the bands excited by the wavepacket. If the medium, where periodic potential is created, is nonlinear, one can observe the formation of lattice solitons that remain localized only due to self-action, whose properties and stability strongly depend on the dimensionality of the problem and that of the potential, see reviews[1-8]. For instance, one-dimensional periodic potentials can support in their bulk stable thresholdless solitons, while in two-dimensional geometries such solitons exist above power threshold. In optics, lattice solitons (or discrete solitons in arrays of weakly coupled waveguides created by shallow refractive index modulations) were predicted and observed in various one-,[9-11] two-,[12-15] and three-dimensional[16-18] settings, see review 4 and references therein. Periodic potentials not only determine the shapes and symmetry of such solitons, but also frequently provide strong stabilizing action, suppressing instabilities that may destroy such self-sustained states in uniform medium. Particularly interesting situation arises when light self-localizes not in the bulk, but in the vicinity of the interface of periodic lattice with uniform medium or with other periodic structure, in which case asymmetric surface lattice solitons may emerge.[19-26] Their characteristic feature is that they also form above power threshold even in one-dimensional geometries. This threshold depends not only on the mean refractive index at both sides of the interface, but also on the details of linear spectra of corresponding materials. Thresholdless surface states appear at specially designed interfaces of two different lattices when edges of some of the forbidden bands of these lattices coincide,[27,28] or at the edges of topologically nontrivial superlattices.[29-32]

Principally new opportunities for control of linear and nonlinear light localization appear when optical lattice is not "static" as in above works, but when it rotates in the direction of light propagation. This is due to the fact that rotation qualitatively changes coupling between lattice sites and this, in turn, dramatically affects diffraction of a wavepacket in such structures. Thus, wavepacket evolution in one-dimensional rotating discrete array mimics dynamics of a quantum harmonic oscillator on a lattice, while in two-dimensional arrays it resembles quantum motion of an electron in the crystalline potential in the presence of a magnetic field.[33] When rotation is applied to the ring-like waveguide arrays with discrete rotational symmetry, it may result in topological suppression of tunneling between azimuthal sites,[34-37] as demonstrated experimentally in refs 38 and 39. In such rotating ring arrays clockwise and counterclockwise vortex currents and vortex solitons become non-equivalent,[40] as shown in experiments with vortices in optically induced structures with several helical channels[41] and in twisted photonic crystal fibers.[42-44] Beyond optics, solitons on rotating ring lattices were studied in Bose-Einstein condensates.[45] Fully two-dimensional rotating periodic arrays can support bulk solitons co-rotating with the array, provided that rotation frequency does not exceed certain critical value,[46-48] above which strong radiation occurs. Nevertheless, first experiments in optically induced extended rotating lattices illustrated specific vector states forming only for high-power excitations shifted from the rotation axis.[49]

Recently, a new mechanism of rotation-induced light localization was theoretically proposed in finite rotating two-dimensional arrays.[50] This mechanism is based on the competition of centrifugal light power transfer and total internal reflection from the borders of finite array and it en-ables the formation of thresholdless two-dimensional solitons (and potentially of robust light bullets[51] in three dimensions), whose localization degree is determined not only by the beam power, but also by the frequency of array rotation. Moreover, rotation allows to create thresholdless solitons even in the center of periodic structure. Our goal is to provide experimental evidence of this new localization mechanism.

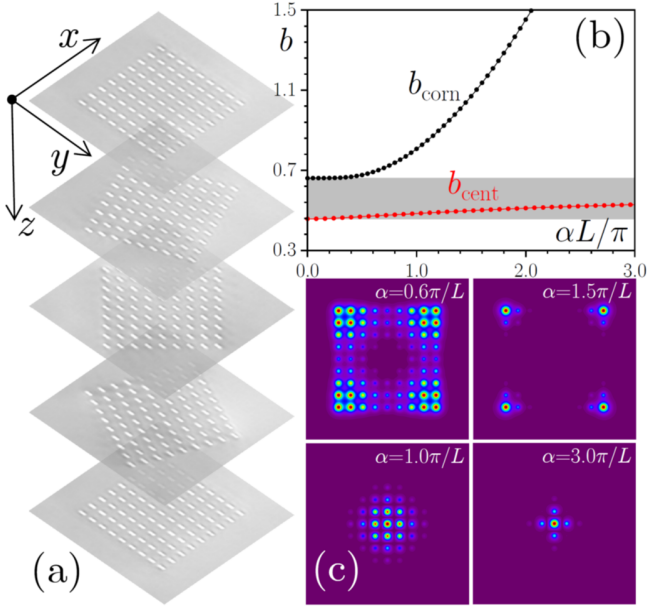

Fig. 1. (a) Stacked microphotographs of the waveguide array at different distances $z$ illustrating its counterclockwise rotation. (b) Eigenvalues of linear modes of the array (black – corner modes, red – central modes) versus rotation frequency $\alpha$ and (c) examples of $|\psi|$ distributions in the corner (top row) and central (bottom row) eigenmodes supported by the rotating array at different $\alpha$. Gray region in (b) shows bulk band.

## RESULTS AND DISCUSSION

In our experiment, rotating square arrays with $9\times9$ waveguides and rotation axis passing through the central waveguide were inscribed in 10 cm-long fused silica glass using the fs-laser writing technique (see Supporting Information for details). Series of counterclockwise rotating arrays with different angular rotation frequencies $\alpha = \phi/L$ were inscribed, where $\phi$ is the angle at which the array rotates at the full sample length $L$. Microscopic images in Fig. 1(a) at different distances $z$ illustrate array rotation with constant $\alpha$. Importantly, our array rotates as a whole around the axis passing through the central waveguide (and on this reason the waveguides located further from the center move along circles of larger radius) in contrast to recently studied Floquet insulators, where each waveguide follows helix of the same radius.[52] Propagation of the light beam in such structures is described by the Schrödinger equation for the dimensionless amplitude of the light field $\psi$, where we also take into account focusing cubic nonlinearity of the material that impacts excitations in the high-power regime:

$$i\frac{\partial \psi}{\partial z} = -\frac{1}{2}\left(\frac{\partial^2 \psi}{\partial x^2} + \frac{\partial^2 \psi}{\partial y^2}\right) - |\psi|^2\psi - \mathcal{R}(x,y,z)\psi, \quad (1)$$

where transverse coordinates $x,y$ are normalized to the characteristic scale $r_0 = 10\ \mu\mathrm{m}$, propagation distance $z$ is normalized to the diffraction length $kr_0^2$ so that the total sample length corresponds to $L\approx 88$, $k = 2\pi n/\lambda$ is the wavenumber at the working $\lambda = 800\ \mathrm{nm}$ wavelength, the function $\mathcal{R} = p\sum_{m,n}\mathcal{Q}(x-x_m, y-y_n)$ describes square array of dimensionless depth $p = k^2 r_0^2 \delta n/n \approx 6.5$ (refractive index contrast $\delta n \sim 7.3\times 10^{-4}$) composed from the identical waveguides $\mathcal{Q}(x,y) = e^{-(x/w_\mathrm{x})^2-(y/w_\mathrm{y})^2}$ with positions of center $x_m = md\cos(\alpha z) - nd\sin(\alpha z)$, $y_n = md\sin(\alpha z) + nd\cos(\alpha z)$ that vary with $z$ due to array rotation, $d = 3.3$ is the period of the array (corresponding to $33\ \mu\mathrm{m}$), the waveguides are elliptical with widths $w_\mathrm{x} = 0.21$ ($2.1\ \mu\mathrm{m}$) and $w_\mathrm{y} = 0.83$ ($8.3\ \mu\mathrm{m}$), their longer axes stay parallel to the $y$-axis of the laboratory coordinate frame at any $z$ due to writing process [Fig. 1(a)].

It is important to stress that despite ellipticity of waveguides leading to a slight anisotropy of the diffraction patterns at small rotation frequencies $\alpha$, due to the fact that orientation of waveguides with respect to primary array axes changes with $z$ upon array rotation (due to writing process), the coupling between them becomes effectively isotropic due to averaging effect of rotation already at frequencies $\alpha \sim \pi/L$. Indeed, by looking at the microphotographs in Fig. 1(a) one can notice that the longer axis of the elliptical waveguides is periodically oriented either along one side of the array or along its perpendicular side and even though this leads to the local increase of coupling in one of the directions in the array, when such "reorientation" occurs several times on 10 cm sample length (for the high enough rotation frequencies), it results in almost uniform diffraction, like if coupling were isotropic in $x$ and $y$ directions.

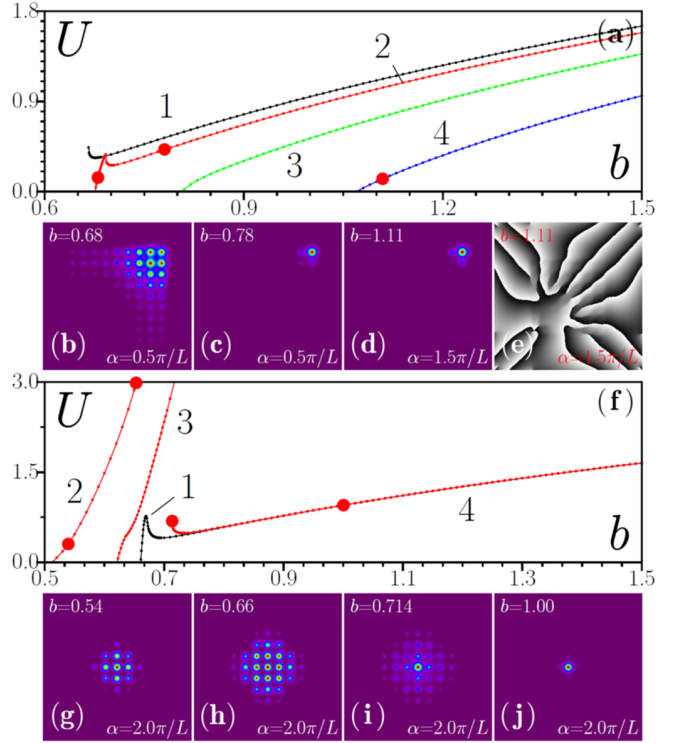

Fig. 2. (a) Power vs propagation constant for corner solitons in arrays with rotation frequency $\alpha = 0.0\pi/L$ (curve 1), $0.5\pi/L$ (2), $1.0\pi/L$ (3), and $1.5\pi/L$ (4). (b)-(e) Examples of $|\psi|$ and nontrivial $\arg(\psi)$ distributions in corner solitons corresponding to the red dots in (a). (f) Power vs propagation constant for solitons in the center of arrays with $\alpha = 0.0\pi/L$ (curve 1) and $\alpha = 2.0\pi/L$ (curves 2,3,4). (g)-(j) Profiles of central solitons corresponding to the red dots in (f) (i.e. shown soliton profiles correspond to curves 2 and 4).

To explain the effect of the rotation-induced localization that occurs even in purely linear regime, we first consider linear low-power excitations and neglect the nonlinear term $|\psi|^2\psi$ in Eq. (1) describing their dynamics. Rotation of the array substantially changes the spectrum of its linear eigenmodes. To illustrate this, we search for the linear eigenmodes $\psi = w(x',y')e^{ibz}$ of the array that remain invariable in the coordinate frame co-rotating with the array: $x' = x\cos(\alpha z) + y\sin(\alpha z)$, $y' = y\cos(\alpha z) - x\sin(\alpha z)$. For simplicity and taking into account the fact that due to rotation the coupling between waveguides becomes effectively isotropic, for this analysis we assume that the waveguides are circular with width $w_{x,y} = 0.36$ (this yields the same effective coupling strength and linear diffraction patterns as for elliptical waveguides already at moderate rotation frequencies). The eigenmodes in the rotating frame can be obtained from the linear eigenvalue problem

$$bw = \frac{1}{2}\left(\frac{\partial^2 w}{\partial x'^2} + \frac{\partial^2 w}{\partial y'^2}\right) - i\alpha\left(x'\frac{\partial w}{\partial y'} - y'\frac{\partial w}{\partial x'}\right) + \mathcal{R}(x',y')w, \quad (2)$$

where representative for rotating systems Coriolis term appears that is proportional to the rotation frequency $\alpha$. It yields localization of some of the modes of the array, whose propagation constants are shown in Fig. 1(b) as functions of the rotation frequency $\alpha$. Among them are corner modes (black dots) forming due to the competition of centrifugal light transfer to the periphery of the array and total internal reflection at the array boundaries that become progressively more localized with increase of $\alpha$ [Fig. 1(c), top]. Four of such modes ($b = b_{\text{corn}}$) bifurcate from the top of the band of delocalized modes of nonrotating array (gray region) and become practically degenerate with increase of $\alpha$. A different type of localized modes (red dots) appears in the center of the array [Fig. 1(c), bottom] due to formation of the attractive "averaged" central optical potential at sufficiently high rotation frequencies (it was shown in refs 47 that in the limit of rapid rotation square array averages into Bessel-like potential that is a function of radius only and that is responsible for appearance of localized mode in the center of the structure). Their propagation constants $b = b_{\text{cent}}$ are located near the bottom of the bulk band (this band is shown here schematically, as it also undergoes complex transformation with increase of $\alpha$). At $\alpha \to 0$ both corner and central linear modes smoothly transform into the delocalized states of nonrotating square array.

Such transformation of linear spectrum also qualitatively impacts the properties of the solitons (see Fig. 2). To obtain them, we now take into account the nonlinear term $|\psi|^2\psi$ in Eq. (1) and search for the soliton solutions in the rotating coordinate frame in the form of $\psi = w(x',y')e^{ibz}$, where the function $w$ satisfies the equation

$$bw = \frac{1}{2}\left(\frac{\partial^2 w}{\partial x'^2} + \frac{\partial^2 w}{\partial y'^2}\right) - i\alpha\left(x'\frac{\partial w}{\partial y'} - y'\frac{\partial w}{\partial x'}\right) + |w|^2 w + \mathcal{R}w, \quad (3)$$

which can be solved using the Newton method. Families of solitons residing in the corner [Fig. 2(a)] or center [Fig. 2(f)] of the rotating array are represented by the dependencies of power $U = \iint |\psi|^2 dx'dy'$ on propagation constant $b$ for different rotation frequencies $\alpha$.

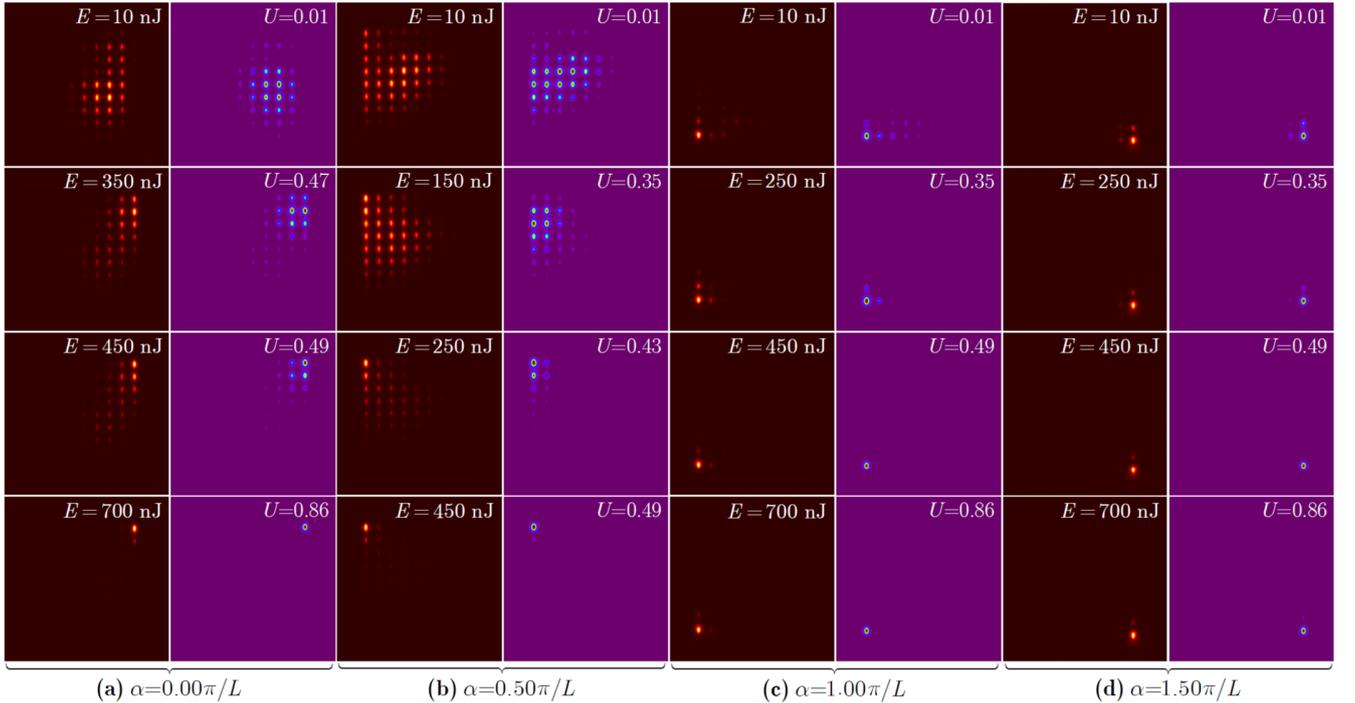

Fig. 3. Comparison of the experimentally measured (left columns, maroon background) and theoretically calculated (right columns, purple background) output intensity distributions for different pulse energies and rotation frequencies for the excitation of upper right corner waveguide in the rotating array. Pulse energies for experimental panels and input powers for theoretical panels are indicated in the top right corners. Rotation frequencies of the array are indicated at the bottom. All intensity distributions are shown within $416~\mu\text{m} \times 416~\mu\text{m}$ windows. Notice that location of the output state (for example, when it remains localized in the corner due to rotation-induced localization) is determined by the total angle at which array rotates at the whole sample length.

At $\alpha = 0$ corner solitons form only above power threshold [Fig. 2(a), curve 1]. Increasing rotation frequency first leads to transformation of the $U(b)$ curve into nonmonotonic dependence (curve 2, notice that the transformation of $U(b)$ dependence with the threshold into the curve without threshold with the gradual increase of $\alpha$ for the corner solitons is not smooth and occurs abruptly around $\alpha \sim 0.3\pi/L$), where maximum of the mode shifts into the bulk of the array when propagation constant approaches the cutoff [Fig. 2(b)]. For sufficiently large $\alpha$ values $U$ monotonically increases with $b$, and solitons become thresholdless since they bifurcate from well-localized linear corner modes [Fig. 2(d)]. In this regime corner solitons are completely stable. All such corner solutions have nontrivial phase distributions reflecting the presence of phase gradient coinciding with the rotation direction [Fig. 2(e)]. It should be mentioned that the stability was tested by propagation of solitons perturbed with the white noise (5% in amplitude) up to the huge distances of $z \sim 10^4$.

For solitons residing in the center of array with $\alpha = 0$ the $U(b)$ curve is nonmonotonic [Fig. 2(f), curve 1], with $U$ vanishing in the cutoff only due to finite size of the array, while in the infinite array such solitons would exist above power threshold. Rotation drastically modifies $U(b)$ dependence leading to its splitting into several branches [we show only several of them here, see red curves 2,3,4 at $\alpha = 2\pi/L$]. This happens because rotation simultaneously localizes several modes in the center of array, with each of them giving rise to the thresholdless soliton family.

The most localized of them is shown by curve 2 [namely this family bifurcates from the red linear state in Fig. 1(b)], but there are other soliton families bifurcating from less localized linear states emerging due to the rotation, like family 3. Interestingly, at fixed $\alpha$ such modes expand with increase of $b$ [see Fig. 2(g),(h)], in contrast to corner solitons that become more localized at higher powers. Families 2 and 3 are stable. Right family 4 of central solitons has different (non-staggered) structure of tails from family 2 and such solitons become more localized with increase of $b$ [Fig. 2(i),(j)]: this family is stable on segments with positive derivative $dU/db$. Notice that for large propagation constant values, central soliton contracts to single waveguide under the action of dominating nonlinear effects and its power becomes practically independent of rotation frequency $\alpha$ - on this reason curves 1 and 4 practically coincide at $b > 1$. It should be also mentioned that family 4 has a threshold because namely linear mode with similar non-staggered phase structure is transformed into the corner state by rotation, so that for this family the nonlinearity that localizes light in central waveguide competes with the effect of rotation.

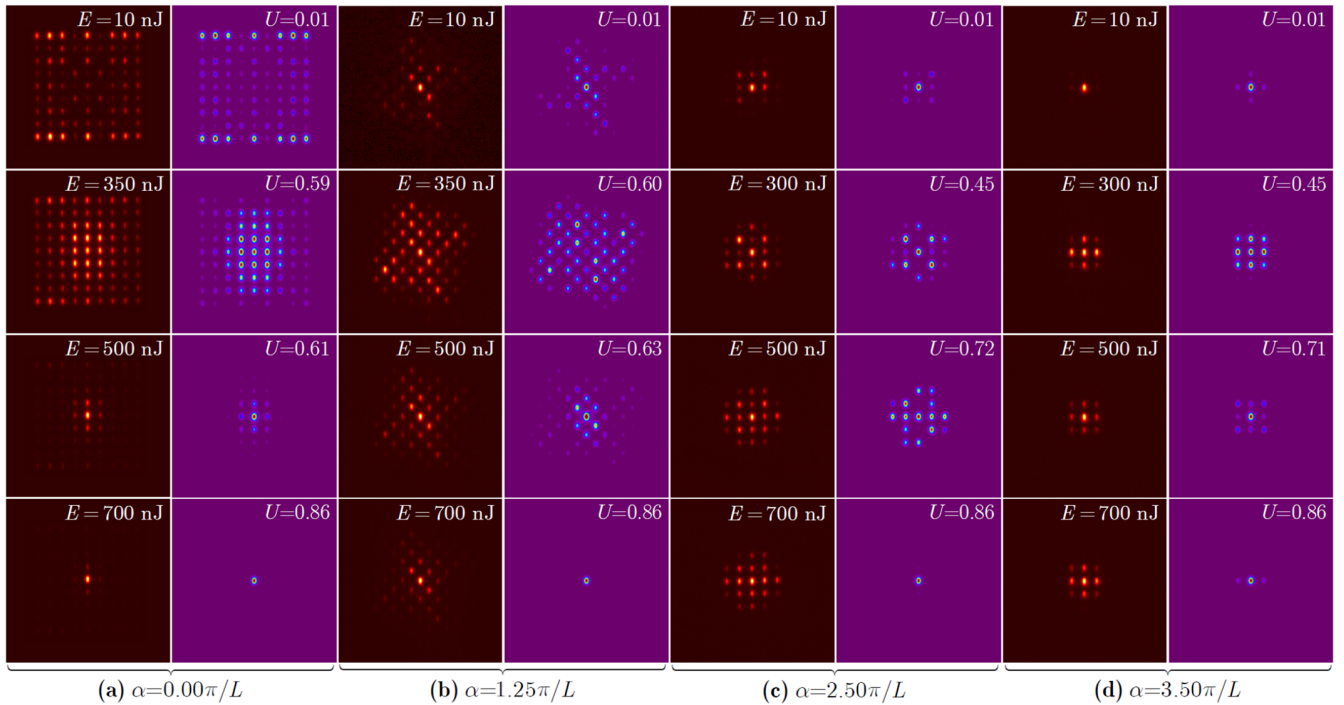

(a) $\alpha=0.00\pi/L$  (b) $\alpha=1.25\pi/L$  (c) $\alpha=2.50\pi/L$  (d) $\alpha=3.50\pi/L$

Fig. 4. Comparison of experimental (left columns, maroon background) and theoretical (right column, purple background) intensity distributions (notations are as in Fig. 3), but for the excitation of central waveguide of the array.

To demonstrate rotation-induced corner localization we excited a single waveguide in the top right corner (other corners yield similar results) of our fabricated structures with different rotation frequencies $\alpha$ using 280 fs pulses of variable energy $E$ from 1 kHz fs Ti:sapphire laser. For details of the experimental setup, excitation of the different waveguides and registration of output intensity distributions see Supporting Information. Figure 3 compares experimental (maroon background) and theoretical (purple background) output intensity distributions at $z=L$. Importantly, in these simulations of dynamical excitation we modeled original Eq. (1) and used elliptical waveguides with longer axis oriented always along the $y$ axis of the laboratory frame, just as in real rotating samples, and Gaussian input beam $\psi|_{z=0} = Ae^{-(x^2+y^2)/\sigma^2}$ with width $\sigma = 0.5$ for the best reproduction of experiments. Thus, these simulations are performed in laboratory frame, where array rotates and therefore the position of the output beam is determined by the total rotation angle of the structure. In nonrotating array [Fig. 3(a)] nonlinear localization in the corner occurs only for sufficiently high pulse energies $E > 450$ nJ. At small rotation frequencies $\alpha \sim 0.50\pi/L$ [Fig. 3(b)] the threshold is clearly reduced and light contracts to the corner already at $E \sim 250$ nJ. For higher frequencies $\alpha = 1.0\pi/L$ [Fig. 3(c)] and $\alpha = 1.5\pi/L$ [Fig. 3(d)] light remains confined in the corner even at the lowest pulse energies, clearly illustrating phenomenon of rotation-induced localization. Notice increasing localization of intensity distributions with increase of pulse energy $E$ for a given $\alpha$ in agreement with the results of Fig. 2(a). To confirm that light remains localized in the corner at all propagation distances, in Fig. S1 of Supporting Information we show that experimental output patterns at the same pulse energies remain practically the same (except for rotation due to rotation of the structure) in $5$ cm and $10$ cm-long samples for a fixed $\alpha = 1.0\pi/L$. Notice that increase of $\alpha$ beyond $2.0\pi/L$ leads to the onset of radiation for corner waveguide.

Experimental evidence of the rotation-induced localization for excitation in the center of array is presented in Fig. 4. In nonrotating array $(\alpha = 0)$ light diffracts strongly in linear regime and transition to localization occurs at higher pulse energies $E > 500$ nJ [Fig. 4(a)] in comparison with corner excitation [Fig. 3(a)]. The tendency for central localization is visible already at rotation frequencies $\alpha \sim 1.25\pi/L$ for low pulse energies [Fig. 4(b)], but with increase of the pulse energy the beam strongly expands and re-localizes only at $E \sim 700$ nJ. Similar nonlinearity-induced expansion of the output pattern is observed at $\alpha = 2.5\pi/L$ [Fig. 4(c)] and $\alpha = 3.5\pi/L$ [Fig. 4(d)]. This delocalization is consistent with the expansion of central solitons with the increase of their power, illustrated for family 2 in Fig. 2(f)-(h). Indeed, when we increase the input power of the Gaussian beam focused into the central waveguide it first excites solitons from family 2 that broaden with the increase of power. However, above the certain power level, one starts exciting predominantly solitons from branch 4 in Fig. 2(f) (the one that has power threshold) and localization of such solitons increases with the increase of power. In dynamical excitation this manifests as initial expansion and then transition to the localization with the increase of input power.

It should be stressed that central solitons excited at low and intermediate pulse energies do maintain their width in the course of rotation, as shown in Fig. S1(c)-(f) comparing outputs in $5$ and $10$ cm samples in Supporting Information, with minimal modifications in their profiles being the result of change of the orientation of elliptical waveguides with respect to primary axes of the array upon its rotation. It should be also stressed that upon re-localization at much higher pulse energies in Fig. 4(c),(d) [connected with more and more efficient excitation of well-localized solitons from family 4 in Fig. 2(f)], one can still observe the background around central peak even at $E \sim 700$ nJ, which is a result of pulsed nature of excitation.

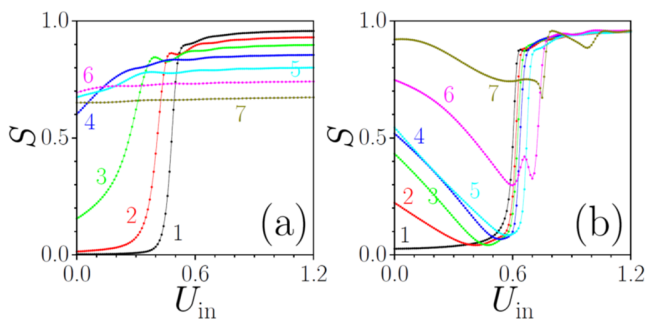

Fig. 5. (a) Power fraction $S$ trapped in corner waveguide vs power of input Gaussian beam $U_{in}$ at $\alpha = 0.0\pi/L$ (1), $0.5\pi/L$ (2), $0.75\pi/L$ (3), $1.0\pi/L$ (4), $1.25\pi/L$ (5), $1.5\pi/L$ (6), and $1.75\pi/L$ (7). (b) $S$ vs $U_{in}$ for excitation of central waveguide at $\alpha = 0.0\pi/L$ (1), $1.0\pi/L$ (2), $1.25\pi/L$ (3), $1.5\pi/L$ (4), $2.0\pi/L$ (5), $2.5\pi/L$ (6), and $3.5\pi/L$ (7).

Finally, in Fig. 5 we present theoretically calculated fraction $S = U_{out}/U_{in}$ of power that remains in the excited corner [Fig. 5(a)] and central [Fig. 5(b)] waveguides versus input power $U_{in}$ (both input $U_{in}$ and output $U_{out}$ powers are calculated here within a ring of radius $2^{1/2}d$ surrounding the excited waveguide). One can see that for corner waveguide with increase of $\alpha$ the excitation efficiency of localized states gradually becomes nearly identical at all power levels, even though it slightly drops down at large $\alpha$ values. For central excitation this dependence clearly illustrates growth of $S$ with $\alpha$ at low powers and delocalization at intermediate power levels (connected with broadening of the central solitons with the increase of $b$), that is replaced by re-localization at high powers.

**CONCLUSIONS**

Summarizing, we demonstrated here a new mechanism of rotation-induced localization of light in sufficiently large rotating waveguide arrays. It enables severe reduction and even elimination of power thresholds for excitation of solitons in corners and in the center of the array. Notice that in arrays of even larger size similar degree of localization for linear corner modes is achieved at the lower rotation frequencies, so that in this sense the impact of rotation in such structures becomes even stronger. At the same time, localization of modes in the center of the structure does not depend on the array size and is determined only by the magnitude of the rotation frequency. Our results pave the way to exploration of rotation-induced localization effects in various photonic systems, like photonic crystals, photonic crystal fibers and multicore fibers. Similar effects can be observed in other physical systems, including cold atoms, Bose-Einstein and polariton condensates. Indeed, the evolution of such systems can be described by similar Schrödinger-like equations, the rotating (with time) truncated optical potentials in them can be created by using the interference patterns of several beams or light modulators, and most of these systems possess nonlinear response allowing to study the localization effects similar to the effects described here, and for various potentials, beyond square ones.

*These authors contributed equally to this work.

**Funding Sources**
1. National Natural Science Foundation of China (NSFC) (11805145)
2. Russian Science Foundation, grant 21-12-00096
3. Research project FFUU-2021-0003 of the Institute of Spectroscopy of the Russian Academy of Sciences
4. China Scholarship Council (CSC) (202006965016)

**Supporting Information**
Supporting Information provides details of the array inscription, details of the experimental setup used for the excitation of different waveguides, and experimental results illustrating steady rotation of corner and central solitons.

# Supporting Information:
# Observation of rotation-induced light localization in waveguide arrays

Chunyan Li,[1,2,*] Antonina A. Arkhipova,[2,3,*] Yaroslav V. Kartashov,[2] Sergey A. Zhuravitskii,[2,4] Nikolay N. Skryabin,[2,4] Ivan V. Dyakonov,[4] Alexander A. Kalinkin,[2,4] Sergey P. Kulik,[4] Victor O. Kompanets,[2] Sergey V. Chekalin,[2] and Victor N. Zadkov[2,3]

[1]School of Physics, Xidian University, Xi'an, 710071, China
[2]Institute of Spectroscopy, Russian Academy of Sciences, 108840, Troitsk, Moscow, Russia
[3]Faculty of Physics, Higher School of Economics, 105066 Moscow, Russia
[4]Quantum Technology Centre, Faculty of Physics, M. V. Lomonosov Moscow State University, 119991, Moscow, Russia

**Fs-laser inscription of rotating waveguide arrays**

Rotating square arrays of $9\times 9$ waveguides with $33~\mu\text{m}$ spacing between them were inscribed in $10~\text{cm}$-long fused silica glass samples (JGS1) using focused (with an aspheric lens with $\text{NA}=0.3$) femtosecond laser pulses at the wavelength $515~\text{nm}$ for pulse duration $280~\text{fs}$, pulse energy $290~\text{nJ}$, and repetition rate $1~\text{MHz}$. Translation of the sample during the writing process of each waveguide was performed by the high-precision air-bearing positioner (Aerotech) with identical for all waveguides velocity of $0.5~\text{mm/s}$. All such waveguides are elliptical ($2.1\times 8.3~\mu\text{m}$), they are single-mode, and they exhibit propagation losses not exceeding $0.3~\text{dB/cm}$ at $\lambda=800~\text{nm}$. A series of square arrays with different angular rotation frequencies $\alpha=\phi/L$ along the $z$-axis, where $L=10~\text{cm}$ is the sample length, and $\phi$ is the total angle by which array rotates at the sample length $L$, were written. In all arrays the axis of rotation passes through the central waveguide of the array. Waveguides are written sequentially, so the waveguide that is on top of the array at $z=0$ during the array rotation goes down and is written "through" the already written waveguides. Despite this, the waveguides are almost identical. Since writing process occurs through the upper surface of the sample, the long axis of the waveguides is always parallel to the $y$-axis of the laboratory coordinate frame, i.e., the orientation of the elliptical waveguides with respect to the primary axes of the array periodically changes with the propagation distance $z$. Due to periodic change of orientation of the elliptical waveguides with respect to array axes, coupling between the waveguides remains on average practically isotropic, as one can see from diffraction patterns in Figs. 3 and 4 of the main text.

**Excitation of corner and central waveguides**

For excitation of corner and central waveguides in rotating array we used light beam from $1~\text{kHz}$ femtosecond Ti:sapphire laser system Spitfire HP (Spectra Physics) delivering $40~\text{fs}$ pulses at $800~\text{nm}$ central wavelength that first pass through an active beam position stabilization system (Avesta) and an attenuator. To prevent optical collapse and strong spectral broadening during propagation in waveguides, short femtosecond pulses with a wide spectrum first enter into a reflective $4\text{f}$ zero-dispersion single-grating stretcher-compressor with a variable slit stretching them to $\tau\sim 280~\text{fs}$ (FWHM) (spectral width was equal to $5~\text{nm}$). Such stretched pulses, whose temporal expansion can be practically neglected on $10~\text{cm}$ sample length were focused into selected waveguides of the array using an optically matched achromatic objective, while output intensity distributions after $10~\text{cm}$ of propagation were registered with the CMOS camera Kiralux (Thorlabs). The input peak power in the waveguide defined as the ratio of the pulse energy $E$ to the pulse duration $\tau$ and taking into account the losses for matching the focusing system can be evaluated as $2.5~\text{kW}$ for each $1~\text{nJ}$.

**Observation of rotating soliton states at different propagation distances**

To prove that the excited states in our rotating waveguide arrays are indeed practically stationary rotating states and not the result of dynamical reshaping, we compared output intensity distributions in arrays with $5~\text{cm}$ and $10~\text{cm}$ lengths for the same pulse energies and for the same rotation frequencies. Such comparison can be performed due to high accuracy of fs-laser inscription technique that guarantees reproducibility of consequently inscribed arrays with the same parameters (period, depth, and rotation frequency). The comparison is presented in Fig. S1(a),(b) for the excitation of the corner waveguide at $\alpha=1.0\pi/L$ and in Fig. S1(c),(d) and Fig. S1(e),(f) for the excitation of the central waveguide in arrays with the rotation frequency $\alpha=2.5\pi/L$ and $3.5\pi/L$, respectively. One can conclude that both central and corner solitons maintain their width in the course of rotation with minimal modifications in their profiles (except for rotation of pattern). Small modifications occur only due to the variation of the orientation of elliptical waveguides with respect to primary axes of the array upon its rotation.

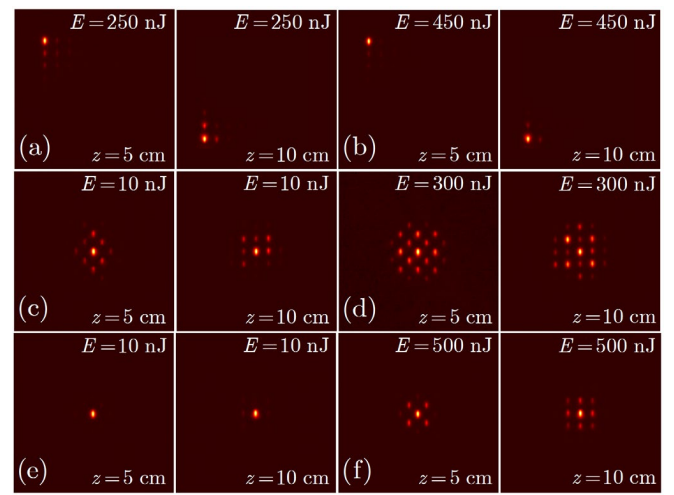

**Fig. S1.** Comparison of nonlinear patterns at a given pulse energy $E$ after $5~\text{cm}$ and $10~\text{cm}$ of propagation in the rotating waveguide array with $\alpha=1.00\pi/L$, corner excitation (a),(b), $\alpha=2.50\pi/L$, central excitation (c),(d), and $\alpha=3.50\pi/L$, central excitation (e),(f).